\documentclass[aps,prl,twocolumn,showpacs,superscriptaddress,floatfix]{revtex4-1}
\pdfoutput=1

\usepackage{amsmath,amssymb}
\usepackage{graphicx}
\usepackage{color}
\usepackage{rotating}
\usepackage{bm}
\usepackage{setspace}
\usepackage{hyperref}
\usepackage{multirow}

\hypersetup{
colorlinks=true,
urlcolor= blue,
citecolor=blue,
linkcolor= blue,
bookmarks=true,
bookmarksopen=false,
}

\renewcommand{\vec}[1]{\bm{#1}}

\begin{document}

%Title of paper
\title{Electric Field Tuning of Band Offsets in Transition Metal Dichalcogenides}

\author{Dennis Huang}
\affiliation{Department of Physics, Harvard University, Cambridge, Massachusetts 02138, USA}
\author{Efthimios Kaxiras}
\email[]{kaxiras@physics.harvard.edu}
\affiliation{Department of Physics, Harvard University, Cambridge, Massachusetts 02138, USA}
\affiliation{John A. Paulson School of Engineering and Applied Sciences, Harvard University, Cambridge, Massachusetts 02138, USA}

\date{\today}

\begin{abstract}
% Present tense
We use first-principles calculations to investigate the band structure evolution of W$X_2$/Mo$X_2$ ($X$ = S, Se) heterobilayers under a perpendicular electric field. We characterize the extent to which the type-II band alignment in these compounds can be tuned or inverted electrostatically. Our results demonstrate two effects of stacking configuration. First, different stackings produce different net dipole moments, resulting in band offset variations that are larger than 0.1 eV. Second, based on symmetry constraints that depend on stacking, a perpendicular electric field may hybridize W$X_2$ and Mo$X_2$ bands that cross at the Brillouin Zone corner $K$. Our results suggest that external electric fields can be used to tune the physics
of intralayer and interlayer excitons in heterobilayers of transition metal dichalcogenides.
\end{abstract}


%\maketitle must follow title, authors, abstract, \pacs, and \keywords
\maketitle

\begin{figure}[t]
\includegraphics[scale=1]{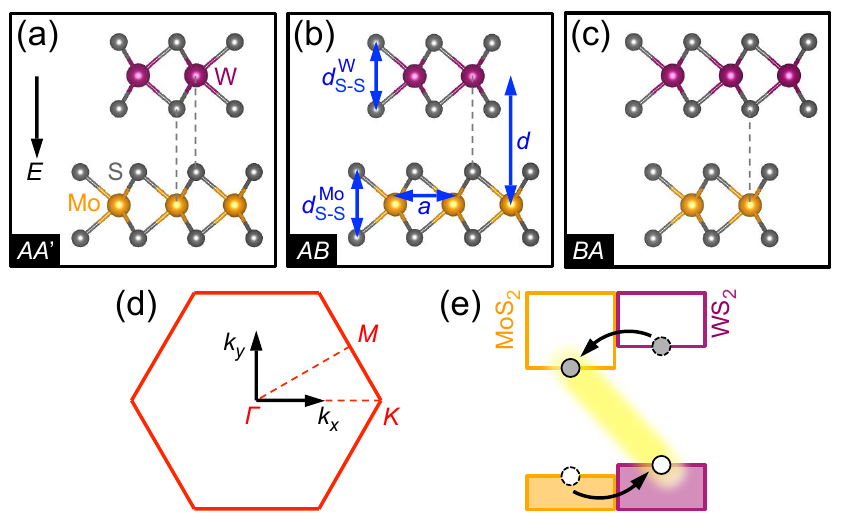}
\caption{Crystal structures of WS$_2$/MoS$_2$ with (a) $AA'$ (180$^{\circ}$), (b) $AB$ (0$^{\circ}$), and (c) $BA$ (0$^{\circ}$) stacking orientations. The black arrow in (a) points in the positive direction of the applied electric field, $E$. (d) Brillouin zone and high-symmetry points corresponding to the $1 \times 1$ unit cell. (e). Schematic of the valence and conduction bands in WS$_2$/MoS$_2$ with staggered, type-II alignment. Upon photoexcitation, the electron and hole are preferentially localized on the MoS$_2$ and WS$_2$ layers, respectively, resulting in an interlayer exciton.}
\label{Fig1}
\end{figure}

% 1. Introduction I
% Present tense
The richness of properties of the layered compounds known as transition metal dichalcogenides (TMDCs) and the ability to stack individual monolayers, either by mechanical transfer or chemical vapor deposition (CVD), afford several degrees of freedom with which to engineer desired functionalities~\cite{Geim_Nat_2013} for a wide range of applications. Single-atomic layers of  TMDCs possess a direct band gap in the visible spectrum~\cite{Mak_PRL_2010, Splendiani_NanoLett_2010} and continue to show promise for many applications that involve electronic excitations, including optoelectronic devices~\cite{Wang_NatNano_2012}. 
% 2. Introduction II
% Past when referring to previous experiments
Of particular interest are heterobilayers of the form W$X_2$/Mo$Y_2$ ($X, Y$ = S, Se), which have staggered, type-II band alignment. In most cases, states at the valence and conduction band extrema are localized on the W$X_2$ and Mo$Y_2$ layers, respectively. Photoexcitation can therefore drive charge separation, wherein hot electrons (holes) migrate to the Mo$Y_2$ (W$X_2$) layer [Fig.~\ref{Fig1}(e)]. Experimentally, ultrafast charge transfer has been observed as a quenching of photoluminescence (PL) signals~\cite{Furchi_NanoLett_2014, Hong_NatNano_2014, Ceballos_ACSNano_2014, Yu_NanoLett_2015, Chiu_NatComm_2015} or broadening of reflectance signals~\cite{Rigosi_NanoLett_2015} from \textit{intralayer} excitons, signaling the dissociation of electron-hole pairs confined to a single layer. Such separation of charge is fundamental to photovoltaic devices~\cite{Furchi_NanoLett_2014, Lee_NatNano_2014}. In many cases, however, the photoexcited electron and hole on the different layers remained bound as an \textit{interlayer} exciton, giving rise to a new PL signal~\cite{Fang_PNAS_2014, Chiu_ACSNano_2014, Gong_NatMat_2014, Rivera_NatComm_2015, Gong_NanoLett_2015, Rivera_Science_2016, He_NanoLett_2016}. Although the detailed properties of interlayer excitons have varied with substrate or annealing conditions~\cite{Tongay_NanoLett_2014, Gong_NatMat_2014, Yu_NanoLett_2015}, the spatial separation of their charges endows them with intrinsic advantages over intralayer excitons. The interlayer excitons have longer radiative~\cite{Rivera_NatComm_2015} and valley~\cite{Rivera_Science_2016} lifetimes, such that their diffusion dynamics can be observed. In addition, the interlayer excitons carry an out-of-plane dipole moment that can be coupled to an external electric field~\cite{Rivera_NatComm_2015}. 

% 3. Introduction III
% Past when referring to previous experiments
Rivera \textit{et al.}~\cite{Rivera_NatComm_2015} demonstrated that the binding energy of interlayer excitons, as measured by PL, undergoes redshifts or blueshifts depending on the direction of an applied electric field. Within back-gated devices of WSe$_2$/MoSe$_2$ and MoSe$_2$/WSe$_2$ on SiO$_2$/Si substrate, they observed shifts up to 0.06 eV. To first order, they attributed this phenomenon to the electrostatic tuning of band offsets. As a caveat, they noted that with a single gate, they could not independently adjust electric field and carrier density, the latter also having an effect through screening~\cite{Chernikov_PRL_2015}. 
Notwithstanding the interpretation of experimental results, it remains crucial to verify theoretically the extent to which an electric field oriented perpendicular to the heterobilayers can modify their band structure features and to elucidate the physics of the induced changes from the fundamental features of the constituents. Such investigations can also provide practical pathways towards realizing excitonic condensates in gate-tunable, double quantum wells of TMDCs~\cite{Fogler_NatCommun_2014}.

% 4. Outline
% Present
In this work, we use density functional theory (DFT) to examine how the band structure of W$X_2$/Mo$X_2$ heterobilayers can be tuned by a perpendicular electric field. Previous theoretical studies of electric field effects have focused on TMDC homostructures~\cite{Ramasubramaniam_PRB_2011, Liu_JPCC_2012, Santos_ACSNano_2013, Zhang_JCP_2014, Xiao_JPCM_2014, Zibouche_PRB_2014, Shanavas_PRB_2015, Nguyen_JEM_2016} and/or band-gap variations~\cite{Lu_Nanoscale_2014, Sharma_JAP_2014}; here, we focus on band offsets at the Brillouin Zone (BZ) corner $K$, where direct optical transitions and valley effects prevail. By constructing different stacking registries, we observe variations in the zero-field band offsets larger than 0.1 eV. Our electric field calculations also reveal stacking-dependent hybridization of W$X_2$ and Mo$X_2$ bands that cross at $K$, which we explain on the basis of wave function symmetries. We discuss potential implications of all these effects on the physics of intralayer and interlayer excitons as probed by PL experiments.

% 5. Methods
% Past
DFT calculations were performed using the Vienna \textit{ab-initio} simulation package (\texttt{VASP})~\cite{Kresse_CMS_1996, Kresse_PRB_1996}, which implements the projector augmented-wave method~\cite{Bloch_PRB_1994, Kresse_PRB_1999}. The valence states are $4s4p5s4d$ for Mo, $5s5p6s5d$ for W, $3s3p$ for S, and $4s4p$ for Se. We used the Perdew-Burke-Ernzerhof (PBE) exchange-correlation functional~\cite{Perdew_PRL_1996}. To simulate a perpendicular electric field, \texttt{VASP} introduces dipole sheets in the middle of the vacuum regions~\cite{Neugebauer_PRB_1992, Makov_PRB_1995}, which we took to be larger than 20 \AA~to avoid interactions between periodic images, making sure that this vaccum size gives well converged energetics and electronic structure features (for details see Supplemental Material). For each heterobilayer and for each value of electric field, we performed an ionic relaxation within the supercell. To obtain accurate interlayer distances, we included van der Waals corrections using rescaled $C_6$ coefficients according to the scheme of Tkatchenko and Scheffler~\cite{Tkatchenko_PRL_2009}, with an energy cutoff of 800 eV. Previous calculations of bulk MoS$_2$ with the Tkatchenko-Scheffler correction produced interlayer distances within 2\% of the experimental value~\cite{Bucko_PRB_2013}. Due to the small dispersion energy, low force tolerances ($<$ 0.2 meV/\AA) were required for convergence. Following the ionic relaxation, we recalculated the non-spin-polarized charge density on a dense grid ($43 \times 43 \times 1$) of the BZ, then used it to obtain the wave function with spin-orbit coupling included.

% 6. Figure 1 and Table 1
% Past when referring to previous experiments
%As shown in Figs.~\ref{Fig1}(a)-(c) and Table~\ref{Tab1}, 
We considered WS$_2$/MoS$_2$ and WSe$_2$/MoSe$_2$ in three stacking configurations, motivated by
experimental and theoretical considerations, labeled 
$AA'$ (180$^{\circ}$), $AB$ (0$^{\circ}$), and $BA$ (0$^{\circ}$), Figs.~\ref{Fig1}(a)-(c),
following the convention of Ref.~\cite{Constantinescu_PRL_2013}: specifically, 
$AA'$ is fully eclipsed; $AB$ is staggered with W atoms aligned with $X$ atoms of the other layer; $BA$ is staggered with Mo atoms aligned with $X$ atoms of the other layer [dashed lines in Figs.~\ref{Fig1}(a)-(c)]. 
Although arbitrary twist angles can be achieved by mechanical transfer, interlayer excitons generated at the $K/K'$ valleys are direct for small angles around 
0$^{\circ}$ and 180$^{\circ}$, resulting in brighter PL~\cite{Yu_PRL_2015}. 
Furthermore, CVD growth of WS$_2$ on MoS$_2$ resulted in $AA'$ stacking~\cite{Gong_NatMat_2014, Yu_NanoLett_2015}, while similar growth 
of WSe$_2$ on MoSe$_2$ resulted in $AA'$ and $AB$ stackings~\cite{Gong_NanoLett_2015, He_NanoLett_2016}. We restricted ourselves to pure chalcogenide compounds ($X$ = $Y$ in W$X_2$/Mo$Y_2$) in order to minimize strain within a commensurate, $1 \times 1$ cell. When optimized, the lattice constants ($a$) of the heterobilayers deviate at most by 0.3\% 
from the lattice constants of their constituent layers in homobilayer form. 
Similarly, the interlayer distances ($d$) of $AA'$-W$X_2$/Mo$X_2$ lie within 0.2\% of the average between $AA'$-bilayer W$X_2$ and $AA'$-bilayer Mo$X_2$. 
At the largest electric field considered (5 V/nm), $d$ values change by 
less than 0.4\% of their zero-field values.

\begin{figure*}[t]
\includegraphics[scale=1]{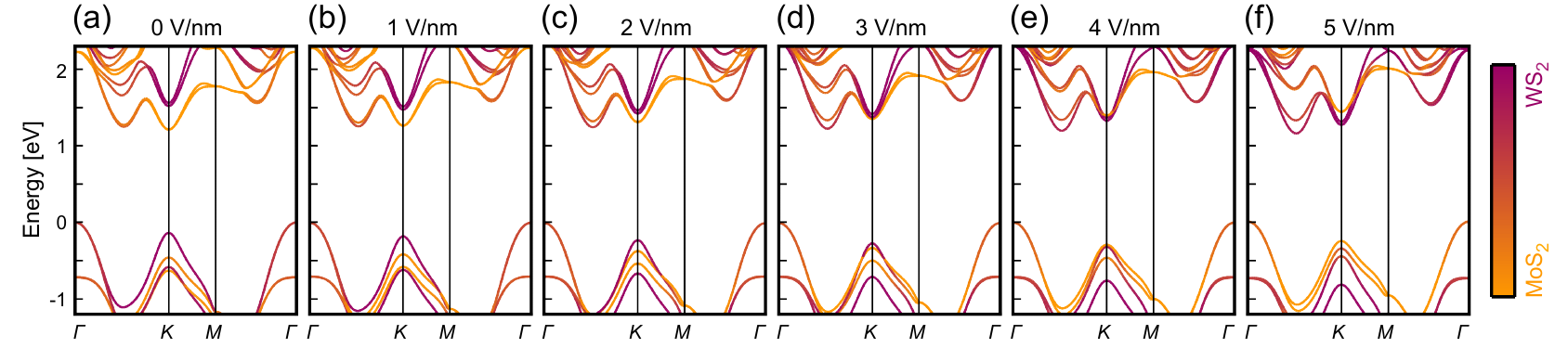}
\caption{(a)-(f) Evolution of the band structure of $AA'$-WS$_2$/MoS$_2$ under a perpendicular electric field. The positive direction of the field points from the WS$_2$ to MoS$_2$ layer [Fig.~\ref{Fig1}(a)]. The zero energy is referenced to the valence band maximum at zero field. The bands are colored based on the projection of their wave functions onto the WS$_2$ and MoS$_2$ layers.}
\label{Fig2}
\end{figure*}

% 7. Figure 2
% Mostly present tense
Figures~\ref{Fig2}(a)-(e) display the band structure evolution of an example heterobilayer, $AA'$-WS$_2$/MoS$_2$, under a perpendicular electric field. High-symmetry points in the BZ are defined in Fig.~\ref{Fig1}(d). At zero field, $AA'$-WS$_2$/MoS$_2$ exhibits an indirect gap of 1.21 eV. The conduction band minimum at $K$ originates from MoS$_2$, whereas the valence band maximum at $\Gamma$ has sizeable contributions from WS$_2$ and MoS$_2$. However, the valence band at $K$ originates from WS$_2$, resulting in a type-II band alignment at $K$ with a direct gap of 1.35 eV. We compare these results with experimental data acquired by scanning tunneling spectroscopy (STS), which probes the single-particle spectrum without excitonic effects~\cite{Ugeda_NatMat_2014}. In a recent measurement of WS$_2$/MoS$_2$ and MoS$_2$/WS$_2$ transferred onto fused quartz, the single-particle gap observed was 1.45 eV~\cite{Hill_NanoLett_2016}. The MoS$_2$ conduction band minimum was found to lie at $K$, whereas it was experimentally uncertain whether the WS$_2$ valence band maximum resided at $K$ or $\Gamma$. 
Previous theoretical works on $AA'$-WS$_2$/MoS$_2$, using hybrid functionals~\cite{Kosmider_PRB_2013} or single-shot $G_0W_0$~\cite{Debbichi_PRB_2014}, predicted gap values of 1.60 eV (direct) and 1.96 eV (indirect) respectively
(none of these calculations included substrate effects, 
which are known to be present in the STS measurements). We note that the order of band alignments predicted by PBE and $G_0W_0$ agree with each other~\cite{Liang_APL_2013}.

% 8. Transition: focus on K
% Mostly present tense
As the electric field increases [in the direction shown in Fig.~\ref{Fig1}(a)], there are various non-rigid transformations in the band structure, such as the shifting of the conduction band minimum from $K$ to a midway point along $\Gamma-K$.
The primary feature of interest for our purposes 
is the band alignment at $K$, which becomes inverted beyond 4 V/nm [Fig.~\ref{Fig2}(e)], as well as the possible hybridization of bands at these crossings. We hereafter focus on the $K$ point, where direct transitions result in bright PL signals~\cite{Yu_PRL_2015} and interlayer excitons exhibit valley-polarized dynamics~\cite{Rivera_Science_2016}.

\begin{figure}[b]
\includegraphics[scale=1]{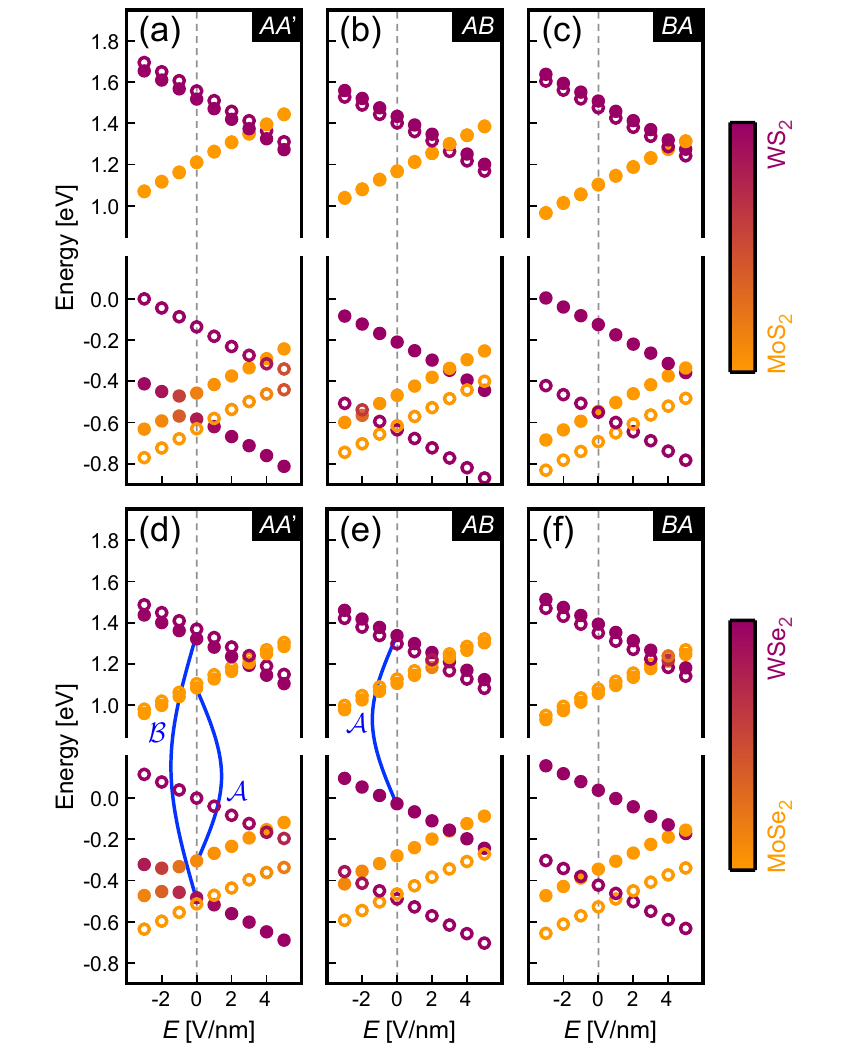}
\caption{Evolution of valence and conduction bands at $K$ under a perpendicular electric field. Three stackings each are shown for (a)-(c) WS$_2$/MoS$_2$ and (d)-(f) WSe$_2$/MoSe$_2$. The zero energies of the WS$_2$/MoS$_2$ and WSe$_2$/MoSe$_2$ compounds are separately referenced to the valence band maximum of their $AA'$ configurations at zero field. The filled (empty) circles denote states with majority up (down) spins, and are colored based on their wave function characters. The curves in (d) and (e) label intralayer $\mathcal{A}$ and $\mathcal{B}$ excitons that are discussed in the main text.}
\label{Fig3}
\end{figure}

% 9. Figure 3
% Present
Figure~\ref{Fig3} shows the evolution of valence and conduction bands at $K$ across a range of positive and negative electric fields, for all six heterobilayers considered. Away from crossings, the energies of W$X_2$-derived bands decrease linearly with electric field, while the energies of Mo$X_2$-derived bands increase linearly with electric field. 
These results show clearly several important trends:\\
(i) Electrostatic inversion of the type-II band alignment is possible, but requires large fields. The $AB$-stacked heterobilayers have the smallest critical fields, which are 3.0 V/nm for WS$_2$/MoS$_2$ and 3.1 V/nm for WSe$_2$/MoSe$_2$, based on linear fits to the spin-up bands. \\
(ii) The magnitudes of the slopes of the band energies versus field decrease in the order of $\textnormal{WS}_2\textnormal{-},$ MoS$_2$-, $\textnormal{WSe}_2\textnormal{-},$ and MoSe$_2$-derived bands. This trend is partly due to the greater polarizability of Se over S.\\
(iii) The distinction between $AB$ and $BA$ stacking being whether the W or Mo atoms are aligned with the $X$ atoms of the opposite layer, these two registries produce net dipole moments at zero field pointing in opposite directions. Thus, to first order, the $BA$ plots are horizontal shifts of the $AB$ plots by 1.7 V/nm for WS$_2$/MoS$_2$ and 1.4 V/nm for WSe$_2$/MoSe$_2$.\\
(iv) Away from crossings, the spin splittings remain relatively constant, and are larger in magnitude for heavier elements. \\
% 10. Figures 4(a), (b)
% Present
(v) There is a sizeable hybridization between valence bands of same spin in the $AA'$ heterobilayers. Smaller hybridizations can also occur when opposite-spin bands cross [for example, in Figs.~\ref{Fig3}(b) and (f)], but they are more difficult to observe due to our discrete sampling of electric field values. 

\begin{figure}[t]
\includegraphics[scale=1]{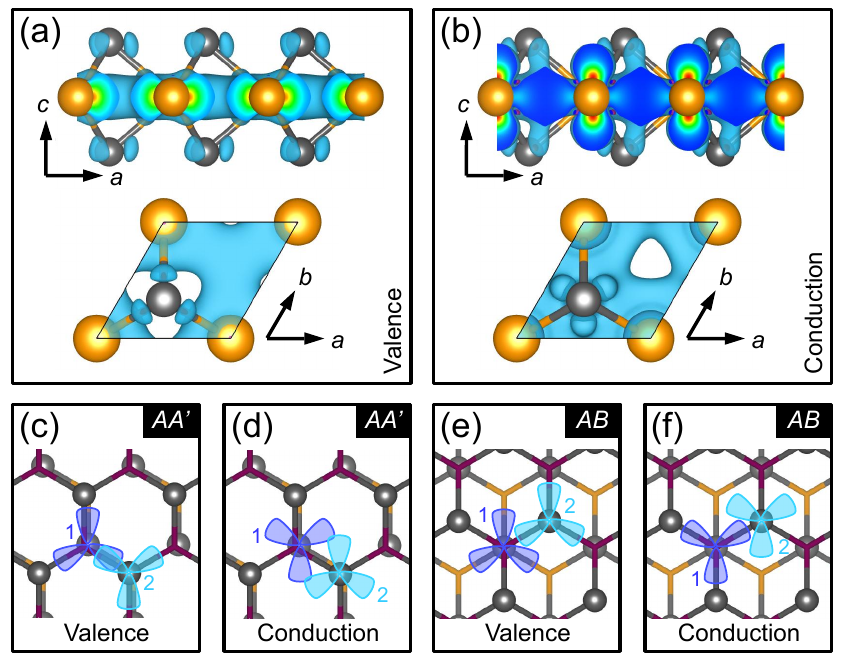}
\caption{(a), (b) Charge density isosurfaces of the valence and conduction band states at $K$ in a monolayer of MoS$_2$. Side and top views are presented for each state. (c)-(f) Schematic depictions of possible interlayer hybridization of $K$ states in $AA'$- and $AB$-stacked WS$_2$/MoS$_2$. Only the S atoms and their charge densities are shown for clarity (1 = MoS$_2$ layer, 2 = WS$_2$ layer). Interlayer hybridization is allowed when the wave functions on both layers transform identically under threefold rotation. The overlap appears largest in (c), where the lobes of charge density are directed towards each other.}
\label{Fig4}
\end{figure}

To elucidate the origin of these effects, we consider the wave functions in the monolayer limit. Figures~\ref{Fig4}(a) and (b) show the charge density isosurfaces of the valence and conduction band states at $K$ in a monolayer of MoS$_2$. Near the S atom, the valence band state has lobes directed along the bonds, whereas the conduction band state has lobes directed between the bonds. Based on Wannier transformations of DFT calculations~\cite{Fang_PRB_2015}, the state at the valence band maximum can be decomposed into $(d_{x^2-y^2} + {\rm i} d_{xy})$-like orbitals on the Mo atom and $(p_x + {\rm i} p_y)$-like orbitals on the S atoms. 
Similarly, the state at the conduction band minimum can be decomposed into $d_{z^2}$-like orbitals on the Mo atom and $(p_x - {\rm i} p_y)$-like orbitals on the S atoms. These states at $K$ have $\mathcal{R}_3$ symmetry (threefold rotation): under a counterclockwise rotation of 120$^{\circ}$ about the S atom, they acquire phase factors of 
$\exp (- {\rm i} 2\pi \mu / 3)$, where $\mu = \pm 1$ for the valence and conduction bands, respectively.

% 11. Figures 4(c)-(f)
% Present
Generalizing to heterobilayers, we note that for $AA'$, $AB$, and $BA$ stacking configurations, $\mathcal{R}_3$ crystalline symmetry is maintained about each atomic site, and is unaffected by a perpendicular electric field. Thus, for $K$ states located on opposite layers to hybridize, their wave functions must transform with the same phase factor under threefold rotation. For a given $K$ state, counterclockwise rotation of 120$^{\circ}$ about a fixed atomic site engenders a phase factor of $\exp ({\rm i} \phi)$, where 
\begin{equation}
\label{Eqphase}
\phi = - \vec{K} \cdot \vec{\delta r} - \frac{2\pi \tau \mu }{3} - \frac{2 \pi s_z}{3}.
\end{equation}
The first term represents a Bloch phase, where $\vec{K} = \frac{4 \pi}{3 a} \hat{x}$ and $\vec{\delta r}$ is a displacement vector of the $X$ atom that may arise from rotation. The second term, which was introduced in the previous paragraph, is generated by orbital angular momentum, with $\tau = \pm 1$ being the valley index. We note that $\tau = -1$ for the 180$^{\circ}$-oriented WX$_2$ layer in the $AA'$ heterobilayers, and $+1$ otherwise. The third term is due to spin, with $s_z = \pm \frac{1}{2}$ for spins pointing in the positive/negative $c$ direction. Using Eq.~\ref{Eqphase}, we derive a set of hybridizations permitted by symmetry, shown in Table~\ref{Tab1}. The actual strength of hybridization depends on matrix elements of the electric field operator, and appears to be greatest in the valence bands of $AA'$ heterobilayers. The stronger hybridization of these states could be due to a greater overlap of the charge densities near the $X$ atoms facing the interlayer cavity [Figs.~\ref{Fig4}(c)-(f)].   

\begin{table}[t]
\center
\setlength{\tabcolsep}{5pt}
\begin{tabular}{c|c|c|c}
\hline 
 & $AA'$ & $AB$ & $BA$ \\
\hline\hline
\multirow{2}{*}{$\mathcal{V}$} & W$X_2$$\uparrow$/Mo$X_2$$\uparrow$ & \multirow{2}{*}{W$X_2$$\downarrow$/Mo$X_2$$\uparrow$} & \multirow{2}{*}{W$X_2$$\uparrow$/Mo$X_2$$\downarrow$} \\
 & W$X_2$$\downarrow$/Mo$X_2$$\downarrow$ & & \\
\hline
$\mathcal{C}$ & W$X_2$$\downarrow$/Mo$X_2$$\uparrow$ & W$X_2$$\downarrow$/Mo$X_2$$\uparrow$ & W$X_2$$\uparrow$/Mo$X_2$$\downarrow$ \\
\hline
\end{tabular}
\caption{Symmetry-allowed hybridizations of valence ($\mathcal{V}$) and conduction ($\mathcal{C}$) bands that cross at $K$. The requisite spin characters from each layer are denoted by the arrows.}
\label{Tab1}
\end{table}

% 12. Discussion I
% Present
In what concerns direct excitons at $K$, our results imply that in the 0$^{\circ}$-stacked heterobilayers, the offset between the spin-up conduction bands defines an \textit{upper bound} on the binding energy of interlayer excitons \textit{relative} to that of the intralayer, W$X_2$ $\mathcal{A}$ exciton~\cite{Rivera_NatComm_2015} [see Fig.~\ref{Fig3}(e)]. The actual binding energy of interlayer excitons is reduced due to the weaker Coulomb interaction between their spatially-separated charges, and vary depending on fabrication details~\cite{Tongay_NanoLett_2014, Gong_NatMat_2014, Yu_NanoLett_2015}. Our results suggest that these energies can be tuned in two ways: 
First, by shifting the stacking registry from $AB$ to $BA$, the spin-up, conduction band offset increases from 0.27 eV to 0.41 eV in WS$_2$/MoS$_2$ and from 0.23 eV to 0.34 eV in WSe$_2$/MoSe$_2$ (DFT total energies for the $AB$ and $BA$ stackings are degenerate within 1 meV, suggesting that the configurations are equally stable). 
Second, by applying electric fields up to $\pm 1$ V/nm, which are attainable in experiments involving dual-gated devices~\cite{Chu_NanoLett_2015}, the spin-up, conduction band offsets vary on average by $\pm 0.09$ eV in WS$_2$/MoS$_2$ and $\pm 0.08$ eV in WSe$_2$/MoSe$_2$.

% 13. Discussion II
% Present
In the 180$^{\circ}$-stacked heterobilayers, the large hybridization of valence bands at $K$ should in principle affect the intralayer, W$X_2$ $\mathcal{B}$ exciton and the intralayer, Mo$X_2$ $\mathcal{A}$ exciton [see Fig.~\ref{Fig3}(d)]. At zero field, the W$X_2$ $\mathcal{B}$ exciton can dissociate due to charge transfer from the spin-up W$X_2$ bands to the spin-up Mo$X_2$ bands, so its PL signal may already be quenched. However, the Mo$X_2$ $\mathcal{A}$ exciton cannot dissociate by hole transfer onto W$X_2$ without a spin flip, which is expected to occur rarely~\cite{Rivera_Science_2016}; thus, its PL signal should be observable. As the system is tuned towards the hybridization point (near $-1$ V/nm for WS$_2$/MoS$_2$), we expect the recombination rate of the Mo$X_2$ $\mathcal{A}$ exciton to drop. This is due to the hole sector of the excitonic wave function becoming spread over both layers, reducing its overlap with the electron sector that is confined to the Mo$X_2$ layer~\cite{Chu_NanoLett_2015}. Thus, valence band hybridization should lead to an observable darkening of the Mo$X_2$ $\mathcal{A}$ exciton.

% 14. Conclusion
In summary, we report 
two effects of the stacking configuration that have impact on optoelectronic properties
of W$X_2$/Mo$X_2$ TMDC heterobilayers:
First, relative shifts in the stacking registry can induce different net dipole moments, leading to band offset variations that affect the binding energy of interlayer excitons. Second, depending on the stacking configuration, a perpendicular electric field may hybridize 
W$X_2$ and Mo$X_2$ bands that cross at $K$, darkening the intralayer excitons associated with these bands. We expect that these stacking effects may be enriched upon introduction of small twist angles about 0$^{\circ}$ or 180$^{\circ}$.

\begin{acknowledgments}
We thank S. Fang and L. A. Jauregui for helpful discussions. Computations were 
performed on the Odyssey cluster supported by the FAS Division of Science, Research Computing Group at Harvard University. Atomic structures and charge density plots were produced using \texttt{VESTA}~\cite{Momma_JAC_2011}. 
We acknowledge support from Army Research Office (ARO-MURI) W911NF-14-1-0247.
\end{acknowledgments}

\newpage

\onecolumngrid

\section*{Supplemental Material}

\setcounter{figure}{0}
\setcounter{equation}{0}
\setcounter{table}{0}
\makeatletter
\renewcommand{\thefigure}{S\@arabic\c@figure}
\renewcommand{\theequation}{S\@arabic\c@equation}
\renewcommand{\thetable}{S\@arabic\c@table}

\subsection{Structural features of heterobilayers}

We summarize the calculated structural features of the relaxed heterobilayers in 
Table \ref{Tab1}. 

\begin{table}[h]
\center
\setlength{\tabcolsep}{6pt}
\begin{tabular}{c|cccccc}
\hline 
 & \multicolumn{3}{c}{\underline{WS$_2$/MoS$_2$}} & \multicolumn{3}{c}{\underline{WSe$_2$/MoSe$_2$}} \\
 & $AA'$ & $AB$ & $BA$ & $AA'$ & $AB$ & $BA$ \\
\hline\hline
$a$ [\AA] & 3.16 & 3.17 & 3.17 & 3.30 & 3.30 & 3.30 \\
$d$ [\AA] & 6.13 & 6.06 & 6.08 & 6.47 & 6.45 & 6.47 \\
$d^{\textrm{Mo}}_{X-X}$ [\AA] & 3.13 & 3.12 & 3.12 & 3.34 & 3.34 & 3.34\\
$d^{\textrm{W}}_{X-X}$ [\AA] & 3.15 & 3.14 & 3.14 & 3.37 & 3.37 & 3.37\\
\hline
\end{tabular}
\caption{Optimized structural parameters [defined in Fig.~\ref{Fig1}(b)] of heterobilayers with various stacking configurations. Values reported are for zero electric field.}
\label{Tab1}
\end{table}

\subsection{\label{sec:A}Van der Waals correction}

\begin{figure}[h]
\includegraphics[scale=1]{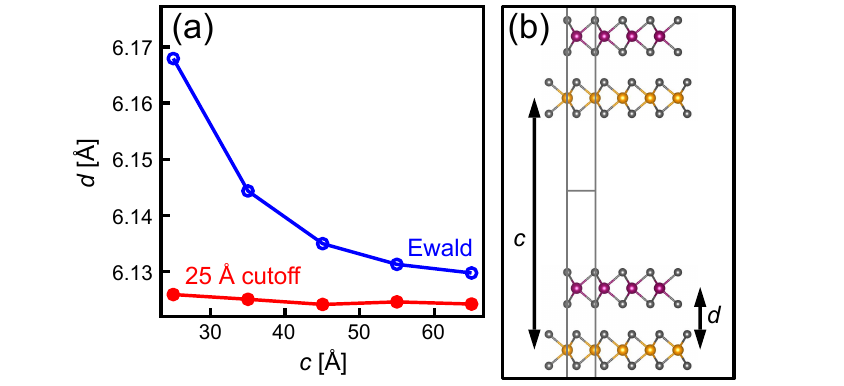}
\caption{(a) Effect of van der Waals correction on the convergence of the interlayer distance ($d$) with respect to the cell length ($c$) in the density functional theory calculation. To obtain convergence, a 25 \AA~cutoff was introduced into the pairwise $r^{-6}$ interactions, in place of a fully periodic summation (Ewald). (b) Structural parameters. We used $c$ = 35 \AA~for the calculations in the main text.}
\label{Fig5}
\end{figure}

% Past tense for actions
Due to the long-range nature ($r^{-6}$) of the empirical van der Waals interaction, an especially large vacuum region within the DFT cell is required to mitigate the influence of periodic images on the TMDC heterobilayer of interest. For a 3D system, Ewald summation of the dispersion energy is both appropriate and accurate~\cite{Kerber_JCC_2008}. However, for our 2D heterobilayer, we found that it slowed the convergence of the interlayer distance ($d$) with respect to the cell dimension along the $c$ axis [Figs.~\ref{Fig5}(a), (b)]. At large values of $c$, the heterobilayer is still slightly pulled apart by its periodic images. To speed up convergence, we truncated the pairwise $r^{-6}$ interactions beyond a cutoff radius of 25 \AA.

\begin{figure}[h]
\includegraphics[scale=1]{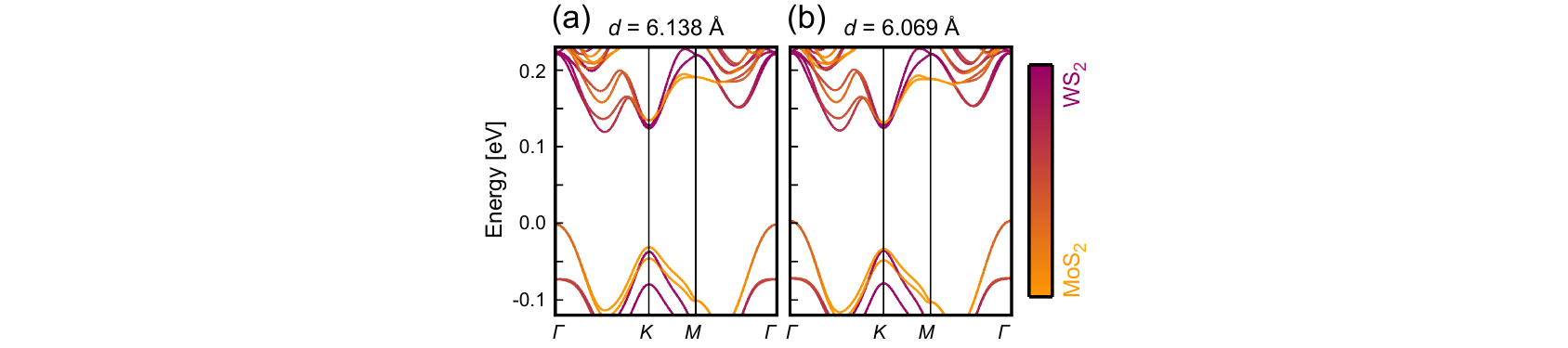}
\caption{Band structure of $BA$-WS$_2$/MoS$_2$ under a perpendicular electric field of 5 V/nm, for interlayer distances $d$ of (a) 6.138 \AA~and (b) 6.069 \AA.}
\label{Fig6}
\end{figure}

Although we strive to calculate the interlayer distance as accurately as possible, small deviations have very little effect on the band structure [Figs.~\ref{Fig6}(a), (b)], and no effect on the hybridization selection rules that are the main focus of our manuscript.

\subsection{\label{sec:B}Artificial vacuum states}

\begin{figure}[h]
\includegraphics[scale=1]{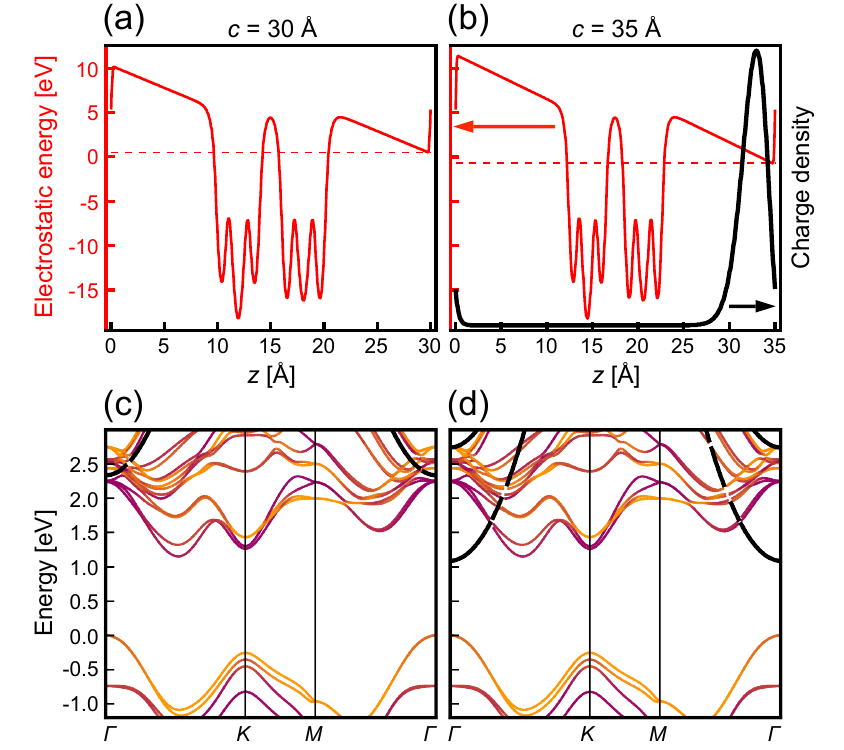}
\caption{(a), (b) Electrostatic potential energy, averaged in the $a$-$b$ plane, for $AA'$-WS$_2$/MoS$_2$ under a field of 5 V/nm. Two calculations are shown, with $c$ = 30 \AA~and 35 \AA. The dashed lines mark the bottom of triangular potential wells induced by periodic boundary conditions. (c), (d) Corresponding band structures for (a), (b). The black lines mark artificial vacuum states trapped by the triangular wells. Their energies are concomitantly lowered with the bottoms of the triangular well as $c$ increases. The charge density corresponding to the lowest vacuum state at $\Gamma$, averaged in the $a$-$b$ plane, is shown in (b).}
\label{Fig7}
\end{figure}

In the implementation of a perpendicular electric field within periodic boundary conditions, a triangular potential well is necessarily created at the boundary of the DFT cell [Figs.~\ref{Fig7}(a), (b)]. This artificial potential can trap quantum well states in the vacuum, leading to the appearance of 2D, free-electron bands centered at $\Gamma$ [Figs.~\ref{Fig7}(c), (d)]. On one hand, the long-range, van der Waals interaction favors enlarging the DFT cell along the $c$ axis, as described previously. On the other hand, for a fixed electric field, the vacuum states get lowered in energy as $c$ increases, leading to difficulties in electronic relaxation if they drop below the valence band maximum. To balance this tradeoff, we took $c$ to be 35 \AA. At this value, the lowest vacuum state at 5 V/nm remains well above the valence band maximum, and the band structure exhibits no change with respect to a calculation with $c$ = 30 \AA. Post calculaton, we removed the vacuum states by dropping bands with negligible projections onto the Mo, W, S, and Se orbitals.

The empty vacuum states may complicate $GW$ calculations, as a large number of bands (up to 1750 in monolayer MoS$_2$~\cite{Fang_PRB_2015}) are typically required for convergence. Further investigations are required to evaluate and possibly negate their influence.

Reference~\cite{Ramasubramaniam_PRB_2011} showed the emergence of a parabolic, $\Gamma$ band in bilayer WS$_2$ at 2.7 V/nm, and Ref.~\cite{Nguyen_JEM_2016} reported a gap closing at the $\Gamma$ point in bilayer MoS$_2$ at 5.5 V/nm. We suggest that these features may originate from artificial vacuum states.

%\bibliography{TMDC_EField}
%merlin.mbs apsrev4-1.bst 2010-07-25 4.21a (PWD, AO, DPC) hacked
%Control: key (0)
%Control: author (72) initials jnrlst
%Control: editor formatted (1) identically to author
%Control: production of article title (-1) disabled
%Control: page (0) single
%Control: year (1) truncated
%Control: production of eprint (0) enabled
%

\end{document}